\documentclass[aps,prb,twocolumn]{revtex4-2}
\usepackage{amsmath,bm}
\usepackage{graphicx,epsfig,braket,amssymb,amsfonts,eufrak,tabularx,multirow,amsmath,graphicx,xcolor}
\usepackage{newtxtext}
\usepackage[colorlinks=true]{hyperref}

\newcommand{\bd}{\begin{displaymath}}
\newcommand{\ed}{\end{displaymath}}
\renewcommand{\v}[1]{{\bf #1}}
\newcommand{\bpm}{\begin{pmatrix}}
\newcommand{\epm}{\end{pmatrix}}
\newcommand{\nn}{\nonumber \\}

\begin{document}

\title{Tunable large spin Nernst effect in a two-dimensional magnetic bilayer}

\author{Gyungchoon Go}
\affiliation{Department of Physics, Korea Advanced Institute of Science and Technology, Daejeon 34141, Korea}

\author{Se Kwon Kim}
\affiliation{Department of Physics, Korea Advanced Institute of Science and Technology, Daejeon 34141, Korea}

\begin{abstract}
We theoretically investigate topological spin transport of the magnon-polarons in bilayer magnet with two-dimensional square lattices.
Our theory is motivated by recent reports on the van der Waals magnets which show the reversible electrical switching of the interlayer magnetic order between antiferromagnetic and  ferromagnetic orders.
The magnetoelastic interaction opens band gaps and allows the interband transition between different excitation states.
In the layered antiferromagnet, due to the interband transition between the magnon-polaron states, the spin Berry curvature which allows the topological spin transport occurs even if the time-reversal symmetry is preserved.
We find that the spin Berry curvature in the layered antiferromagnet is very large due to the small energy spacing between two magnon-like states.
As a result, the spin Nernst conductivity shows sudden increase (or decrease) at the phase transition point between ferromagnetic and antiferromagnetic phases.
Our result suggest that the ubiquity of tunable topological spin transports in two dimensional magnetic systems.
\end{abstract}

\maketitle

\section{Introduction}\label{sec1}

Magnonics and phononics are evolving research field in modern condensed matter physics.
In magnetic insulator, the energy and information are carried by the magnetic excitations (magnons) and lattice vibrations (phonons).
Because these collective low-energy excitations open up the possibility of new schemes to manipulate and control the thermal energy and information,
intensive studies have focused on these charge neutral quasi-particles~\cite{Chumak2015,Maldovan2013}.
On this issue, one of the fundamental interest is the topological Hall transport of the collective excitations due to the Berry curvature.
Previous reports demonstrated that the magnonic and phononic systems can have the topological bands exhibiting
the magnon Hall effect in chiral magnetic systems~\cite{Katsura2010, Onose2010, Matsumoto2011a, Matsumoto2011b, Shindou2013a, Shindou2013b, Zhang2013, Mook2014, Kimsk2016, Cheng2016, Owerre2016, Zyuzin2016, Nakata2017, Gobel2017, Kimsk2019, Diaz2019} and the phonon Hall effect in the presence of the spin-phonon interaction~\cite{Strohm2005, Sheng2006, Kagan2008, Inyushkin2007, LZhang2010, Ideue2017, Sugii2017}, respectively.

Recent years, another mechanism for the topological Hall transport of the quasi-particles has been proposed.
The magnon-phonon interaction allows the nontrivial topology of the hybrid excitation of the magnon and phonon (called magnon-polaron)
even though each of magnonic and phononic system is topologically trivial~\cite{Takahashi2016, Park2019, XZhang2019, Go2019, SZhang2020}.
The magnon-polaron bands can have the nontrivial topology in the presence of the long-range dipolar interaction~\cite{Takahashi2016},
the Dzyaloshinskii-Moriya interaction~\cite{XZhang2019}, or the exchange-induced magnetoelastic coupling in noncollinear antiferromagnet (AFM)~\cite{Park2019}.
It has been also shown that the topological magnon-polaron can be obtained in a simple square lattice collinear ferromagnet (FM)~\cite{Go2019} and AFM~\cite{SZhang2020}
by the anisotropy-induced magnetoelastic coupling~\cite{Kittel1949}.
The topology of the magnon-polaron results in the thermal Hall transport of the quasi-particles.
For the topological spin transport of the magnon-polarons, the Dzyaloshinskii-Moriya-interaction-induced spin-Nernst effect has been investigated~\cite{Bazazzadeh2021}.
However, the pure spin transport carried by the topological magnon-polarons without the Dzyaloshinskii-Moriya interaction is still unexplored.

Meanwhile, intrinsic magnetism in two-dimensional (2D) van der Waals materials have been discovered recently
and attracted growing attention because of the fundamental scientific interest and unprecedented oppotunities for technological applications in reduced dimensions~\cite{McGuire2015, Zhang2015, Lee2016, Gong2017, Huang2017, Bonilla2018, OHara2018, Fei2018, Deng2018, Burch2018, Gibertini2019}.
In particular, the chromium triiodide (CrI$_3$) has a robust magnetism in monolayer limit with a strong perpendicular magnetic anisotropy~\cite{McGuire2015, Huang2017}.
The monolayer CrI$_3$ is a ferromagnet (FM) while a pristine bilayer CrI$_3$ is an antiferromagnet (AFM) composed of two FM monolayers coupled antiferromagnetically~\cite{Huang2017}.
Because the topological particle transports of the magnon-polarons are allowed in monolayer FM,
the AFM phase of the bilayer CrI$_3$ is expected to show the pure spin transport of the topological magnon-polarons from its time-reversal symmetry.
Moreover, recent experiments demonstrate that an electric gating can change the AFM interlayer order in the bilayer CrI$_3$ to FM order and vice versa~\cite{Jiang2018a, Jiang2018b, Huang2018}
which enables the electric field control of the topological quasi-particle and the spin transport.

\begin{figure}[t]
\includegraphics[width=86mm]{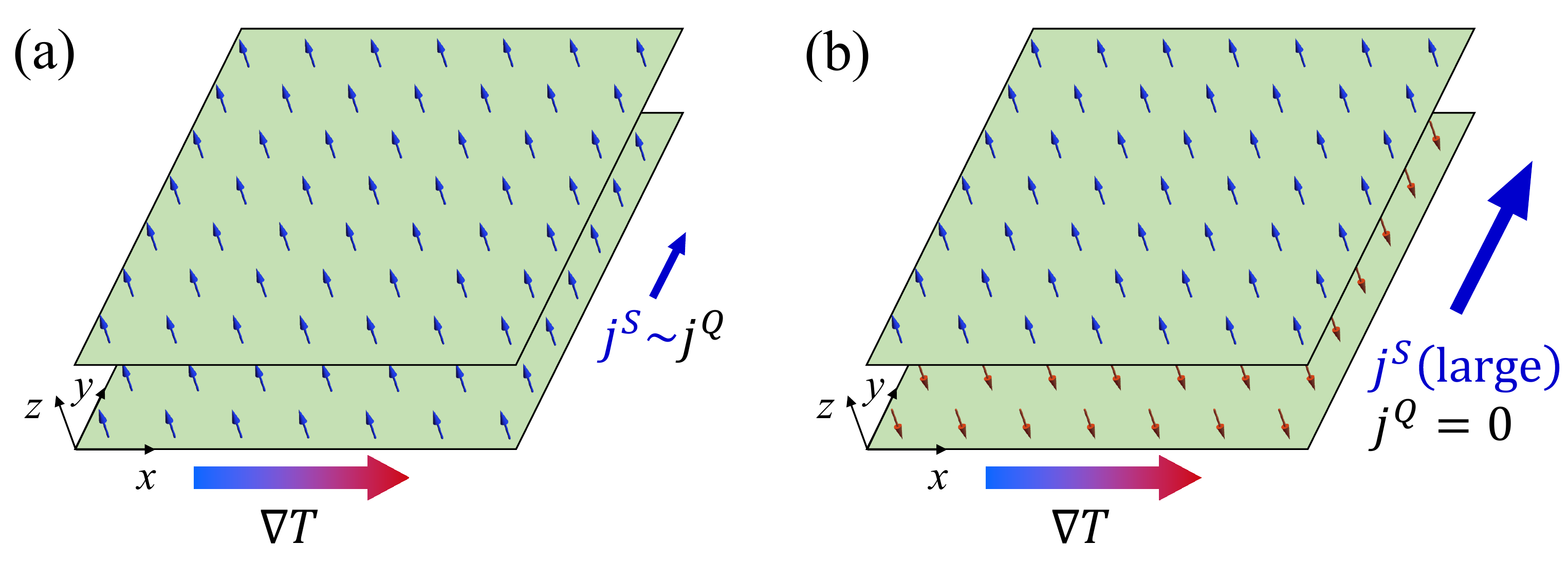}
\caption{Schematic illustration of the thermal Hall and spin Nernst effects in the bilayer (a) ferromagnet (FM) and (b) antiferromagnet (AFM).
(a) In a layered FM, the topological magnon-polarons exhibit the Hall response under a temperature gradient.
Because the magnon possesses the spin angular momentum, the thermal Hall transport of the magnon-polaron ($j^Q$) leads to the transverse spin current ($j^S$).
(b) In a layered AFM, the thermal Hall effect of the magnon-polaron is prohibited by the time-reversal symmetry
whereas the spin Nernst effect is induced by the magnetoelastic interaction which is shown to be large compared to that in the layered FM (see the main text).
} \label{fig:1}
\end{figure}

In this paper, we theoretically investigate the topological spin transport of the magnon-polarons in a bilayer system of square-lattice FM layers [see Fig.~\ref{fig:1}(a)].
We find that the magnetoelastic interaction in the layered AFM induces the finite spin Berry curvature to the magnon-polaron bands in the absence of the magnetic field.
Different from previous studies~\cite{Go2019,SZhang2020}, the dominant contribution of the topological spin transport does not originate from the band-anticrossings
between a magnonic and a phononic bands.
In the AFM, the magnetoelastic interaction opens band gaps between two magnon-like bands which degenerate in the absence of the magnetoelastic interaction.
Because these band gaps are much smaller than the anticrossing gap between a magnonic and a phononic bands,
the spin Berry curvature from this effect is large when the level broadening effect is small.
We also compute the spin Nernst and thermal Hall conductivities in the layered magnet
and find that a large enhancement of the spin Nernst conductivity when the interlayer exchange becomes antiferromagnetic.

The remaining part of this paper is organized as follows.
In Sec.~\ref{sec2}, we consider a minimal model describing the magnon-polarons in the bilayer AFM.
In this section, we provide a low-energy effective Hamiltonian and compute the spin Berry curvature of the magnon-polaron bands.
In Sec.~\ref{sec3}, we compute the thermal Hall and spin Nernst effects of the magnon-polarons.
In Sec.~\ref{sec4}, we conclude with a brief summary and discussion.

\section{Topological magnon-polaron in layered antiferromagnet}\label{sec2}

\subsection{Model Hamiltonian}

We start from a bilayer antiferromagnet (AFM) on the 2D square lattice described by the Hamiltonian
\begin{align}
H = H_{m} + H_{ph} + H_{mp},
\end{align}
where the magnetic part is
\begin{equation}
\begin{aligned}
H_{m} =& -J \sum_{l \in \{A,B\}}\sum_{\langle i,j\rangle} \v S^l_i\cdot \v S^l_j - J_{AB} \sum_i \v S^A_i\cdot \v S^B_i\\
&- \frac{K_z}{2} \sum_{i,l} (S^l_{i,z})^2 - {\cal B} \sum_{i,l} S^l_{i,z},
\end{aligned}
\end{equation}
where ${\langle i,j\rangle}$ runs over all pairs of nearest neighbors, $J > 0$ is the ferromagnetic Heisenberg exchange interaction, $K_z >0 $ is the perpendicular easy-axis anisotropy, and $\cal B$ is the external magnetic field along the easy-axis, $J_{AB}$ is the Heisenberg exchange interaction between upper ($l = A$) and lower ($l = B$) layers.
Because $J > 0$, each layer is a ferromagnet in which the ground state is the uniform spin configuration along the easy-axis.
When the interlayer exchange coupling is negative ($J_{AB} <0$), the ground state spin configurations of A and B layers are opposite to each other ($i.e.,$ $\v S_{A} = S \hat {\v z}$ and $\v S_{B} = -S \hat {\v z}$). The magnon Hamiltonian of the layered AFM can be obtained by performing the Holstein-Primakoff transoformation with the Bogoliubov transformation
\begin{align}
H_m =&\sum_{\v k} \left [\epsilon^{\alpha}_{m,\v k} \alpha^\dag_{\v k} \alpha_{\v k} + \epsilon^{\beta}_{m,\v k} \beta^\dag_{\v k} \beta_{\v k}\right],
\end{align}
where the eigenenergy of the $\alpha (\beta)$ mode is
\begin{align}\label{Hmag}
\epsilon^{\alpha/\beta}_{m,\v k} = \sqrt{(\epsilon^0_{m,\v k})^2 + 2 \epsilon^0_{m,\v k}|J_{\rm AB}| S} \pm {\cal B},
\end{align}
and $\epsilon^0_{m,\v k} = 2JS\left(2-\cos{k_x} - \cos{k_y}\right) + K_zS$ is the monolayer magnon dispersion without the external magnetic field.
We note that the two magnonic modes ($\alpha$- and $\beta$-modes) degenerate without the  magnetic field.

For the phonon Hamiltonian describing the lattice dynamics, we focus on the
out-of-plane components of the displacement vector which is the only component that gives the nontrivial topology through the magnetoelastic coupling~\cite{Go2019,SZhang2020}. Then, we have
\begin{equation}
\begin{aligned}\label{Hph}
H_{ph} = &\sum_{i,l} \frac{(p_{z,i}^{l})^2}{2M} + \frac{\lambda}{2} \sum_{\langle i,j\rangle,l} (u^l_{z,i} - u_{z,j}^l)^2\\
&+ \frac{\lambda_{AB}}{2} \sum_i (u^A_{z,i} - u_{z,j}^B)^2,
\end{aligned}
\end{equation}
where ${u}^l_{z,i}$ and $p^l_{z,i}$ are the out-of-plane displacement of the $i$th ion in the $l$th layer and its conjugate momentum, respectively,
$M$ is the ion mass, and $\lambda$ is the spring constant between nearest-neighbors in the same layer, $\lambda_{AB}$ is the interlayer spring constant.
Because of the interlayer lattice coupling $\lambda_{AB}$, the lattice vibration modes split into the anti-bonding and bonding modes.
By quantizing the lattice displacements and its conjugate momenta
\begin{equation}\label{Equp}
\begin{aligned}
&u^\eta_{z,\v k} = \sqrt{\frac{\hbar}{M {\omega}^\eta_{p,\v k}}} \left(\frac{b_{\eta,\v k} + b^\dag_{\eta,-\v k}}{\sqrt2}\right),\\
&p^\eta_{z,\v k} = \sqrt{\hbar M {\omega}^\eta_{p,\v k}}  \left(\frac{b_{\eta,-\v k} - b^\dag_{\eta,\v k}}{{\sqrt2} i}\right),
\end{aligned}
\end{equation}
we have the corresponding phonon Hamiltonian
\begin{align}\label{Hphonon}
H_{ph} &= \sum_{\eta =1}^2 \sum_{\v k} \hbar \omega^\eta_{p,\v k} b^{\dag}_{\eta,\v k} b_{\eta,\v k},
\end{align}
where $\omega^1_{p,\v k} = \sqrt{(\omega_{p,\v k})^2 + 2\omega_z^2}$ is the dispersion of the anti-bonding phonon
and $\omega^2_{p,\v k} = \omega_{p,\v k}$ is the dispersion of the bonding phonon.
Here, $\omega_{p,\v k} = \omega_{0} \sqrt{(4 - 2 \cos k_x - 2\cos k_y)}$ is the monolayer phonon frequency.
The characteristic frequencies are $\omega_0 = \sqrt{\lambda/M}$ and $\omega_{z} = \sqrt{\lambda_{AB}/M}$.
Because we are dealing with the out-of-plane phonon, the interlayer phonon coupling $\omega_z$ is large when the interlayer distance is sufficiently small.
In this case, the energy of the anti-bonding phonon ($\hbar \omega^1_{p,\v k}$) is much higher than the other bands.
Because our interest is on the low-energy excitations, we neglect the anti-bonding phonon mode and
drop the index $\eta\,$ ($i.e$., $b_{\v k} \equiv b_{2,\v k}$).

For the magnon-phonon interaction, we adopt the elastic strain derived by Kittel~\cite{Kittel1949, Kittel1958}.
To linear order in the magnon amplitude, this is given by~\cite{Thingstad2019}
\begin{align}\label{mpint}
H_{mp} = \kappa \sum_{i,l} \sum_{\v e_i} \left(\v S^l_i \cdot \v e_i \right) \left(u^l_{z,i} - u^l_{z,{i+\v e_i}} \right),
\end{align}
where $\kappa$ is the strength of the magnetoelastic coupling and $\v e_i$ are the nearest neighbor vectors.
We note that the magnon-phonon interaction contains both particle-number-conserving terms and particle-number-nonconserving terms.
If we consider $H_{mp}$ as a weak perturbation with well defined energies of magnons and phonons,
the magnetoelastic Hamiltonian expressed in magnon and phonon operators is dominated by particle-number-conserving terms near the band-crossing points~\cite{Go2019, SZhang2020}.
Neglecting the particle-number-nonconserving terms, we have
\begin{equation}\label{Hmp}
\begin{aligned}
H_{mp} =i\sum_{\v k}  {\bar\kappa}\left( s_-\alpha^\dag_{\v k} b_{\v k}  + s_+ \beta^\dag_{\v k} b_{\v k}\right) + h.c.,
\end{aligned}
\end{equation}
where ${\bar \kappa} = \kappa \sqrt{\frac{\hbar S\sqrt{(1+2\Delta)}}{2M {\omega}_{p,\v k}}}$, $\Delta = {|J_{\rm AB}| S}/{\epsilon^0_{m,\v k}}$, and
$s_{\pm} = \sin k_x \pm i \sin k_y$.

From Eq.~\eqref{Hmag}, ~\eqref{Hphonon}, and ~\eqref{Hmp}, we write an effective three-band Hamiltonian
\begin{align}\label{simpHk}
H_{\rm eff} \approx \sum_{\v k} \psi^\dag_{\v k} \left(
                                     \begin{array}{cccc}
                                       \epsilon^\alpha_{m,\v k} & 0 & i {\bar\kappa }s_-  \\
                                       0 & \epsilon^\beta_{m,\v k} & i{\bar\kappa }s_+  \\
                                       -i{\bar\kappa } s_+ & -i{\bar\kappa } s_- & {\epsilon}_{p, \v k} &  \\
                                     \end{array}
                                   \right)
\psi_{\v k},
\end{align}
where $\psi_{\v k} = (\alpha_{\v k}, \beta_{\v k}, b_{\v k})$ and ${\epsilon}_{p, \v k} = \hbar \omega_{p,\v k}$.
In the presence of the time-reversal symmetry (${\cal B} = 0$),
$\alpha$- and $\beta$-magnons are energetically degenerate without the magnon-phonon interaction ($\epsilon^\alpha_{m,\v k} = \epsilon^\beta_{m,\v k} \equiv \epsilon_{m,\v k}$).
In this case, the energy eigenvalues of the magnon-polaron Hamiltonian~\eqref{simpHk} are simplified to
\begin{equation}
\begin{aligned}
\epsilon_{1,\v k} &= \epsilon^+_{\v k} - \sqrt{(\epsilon^-_{\v k})^2 + \bar{\kappa}^2 f(\v k)},\\
\epsilon_{2,\v k} &= \epsilon_{m,\v k},\\
\epsilon_{3,\v k} &= \epsilon^+_{\v k} + \sqrt{(\epsilon^-_{\v k})^2 + \bar{\kappa}^2 f(\v k)},
\end{aligned}
\end{equation}
where $\epsilon^\pm_{\v k} = (\epsilon_{m, \v k} \pm \epsilon_{p, \v k})/2$ and $f(\v k) = 2 - \cos(2 k_x) - \cos(2 k_y)$.
In Fig.~\ref{fig:2}(a), we show the bulk band structure without external magnetic field.
Here we adopt following parameters used in Ref.~\cite{Go2019}:
$J= 2.2$ meV, $K_z = 1.36$ meV, $S = 3/2$, $M c^2 = 5\times 10^{10}$ eV, $\hbar \omega_0 = 10$ meV, $\kappa = 5$ ${\rm meV}/{\rm \mathring{A}}$.
For the interlayer couplings, we assume that $\omega_z = 2 \omega_0$ and $J_{\rm AB} = -J/2$.
The magnon-phonon interaction hybridizes the magnon and phonon bands and opens the band gap between $\epsilon_1 (\v k)$ and $\epsilon_3 (\v k)$.
However, $\epsilon_2 (\v k)$ is not perturbed by the magnon-phonon interaction because the corresponding eigenstate is a mixture of two magnon states ($\alpha$- and $\beta$- magnons) without the phonon (see Appendix~\ref{secA1}).

\subsection{Topological magnon-polarons}

Because the $\alpha$- and $\beta$-magnons carry opposite spin angular momenta
($\langle\alpha_\v k |S_z|\alpha_\v k \rangle = -1$, $\langle\beta_\v k|S_z|\beta_\v k \rangle = +1$),
the three-band Hamiltonian~\eqref{simpHk} can lead to the topological spin transport even if the time-reversal symmetry is preserved.
The topological spin transport of the $n$-th band is characterized by the spin Berry curvature (sometimes called generalized Berry curvature)~\cite{Li2020,Qu2021,Bazazzadeh2021}
\begin{align}
\Omega_{xy}^{{s,n}}(\v k) =  \lim_{\tau \rightarrow \infty} \Omega_{xy}^{{s,n}}(\v k,\tau),
\end{align}
where
\begin{align}
\Omega_{xy}^{{s,n}}(\v k,\tau) = \sum_{m\neq n} \frac{{\rm Im} \left[-2\hbar^2 \langle n_\v k| j^{s,z}_x |m_\v k\rangle
\langle m_{\v k}| v_y |n_{\v k}\rangle\right]}{(\epsilon_{n,\v k} - \epsilon_{m,\v k})^2  + \left(\frac{\hbar}{\tau}\right)^2}.
\end{align}
where $v_i = \frac{1}{\hbar}\frac{\partial {\cal H}_{\v k}}{\partial k_i}$ and $j^{s,z}_x = \frac{1}{2}\{S_z,v_x\}$ represent the velocity and the spin current operators, respectively.
Here, $|n_\v k\rangle$ and $\epsilon_{n,\v k}$ are the eigenstate and eigenenergy of $n$-th band, respectively.
The relaxation time $\tau \equiv (1/\tau_n + 1/\tau_m)^{-1}$ is introduced to capture the level broadening effect for thermal transport in next section.
For better understanding, we seperate the spin Berry curvature into two components
\begin{align}\label{SSB}
\Omega_{xy}^{{s,n}}(\v k) = \Omega_{xy,\rm I}^{{s,n}}(\v k) + \Omega_{xy,\rm II}^{{s,n}}(\v k).
\end{align}
The type-I spin Berry curvature is written as
\begin{align}\label{SBEI}
\Omega_{xy,\rm I}^{{s,n}}(\v k) &=  \sum_{m\neq n} S_{z,\v k}^{n}\frac{- \hbar^2{\rm Im}  \left[\langle n_{\v k}|v_x |m_\v k\rangle\langle m_{\v k}| v_y |n_{\v k}\rangle\right]}{(\epsilon_{n,\v k} - \epsilon_{m,\v k})^2 }\nn
&\hspace{11mm}- (n \leftrightarrow m)\nn
&= \sum_{m\neq n} \left(S_{z,\v k}^{n} + S_{z,\v k}^{m}\right)\Omega_{xy}^{{nm}}(\v k),
\end{align}
where $S_{z,\v k}^{n} = \langle n_{\v k}| S_z |n_{\v k}\rangle$ is the spin expectation value of the $n$-th band.
This term describes spin Hall transport due to the (projected) Berry curvature $\Omega_{xy}^{{nm}}(\v k)$ which vanishes when the time-reversal symmetry is preserved.
The type-II spin Berry curvature is
\begin{widetext}

\begin{figure}[t]
\includegraphics[width=130mm]{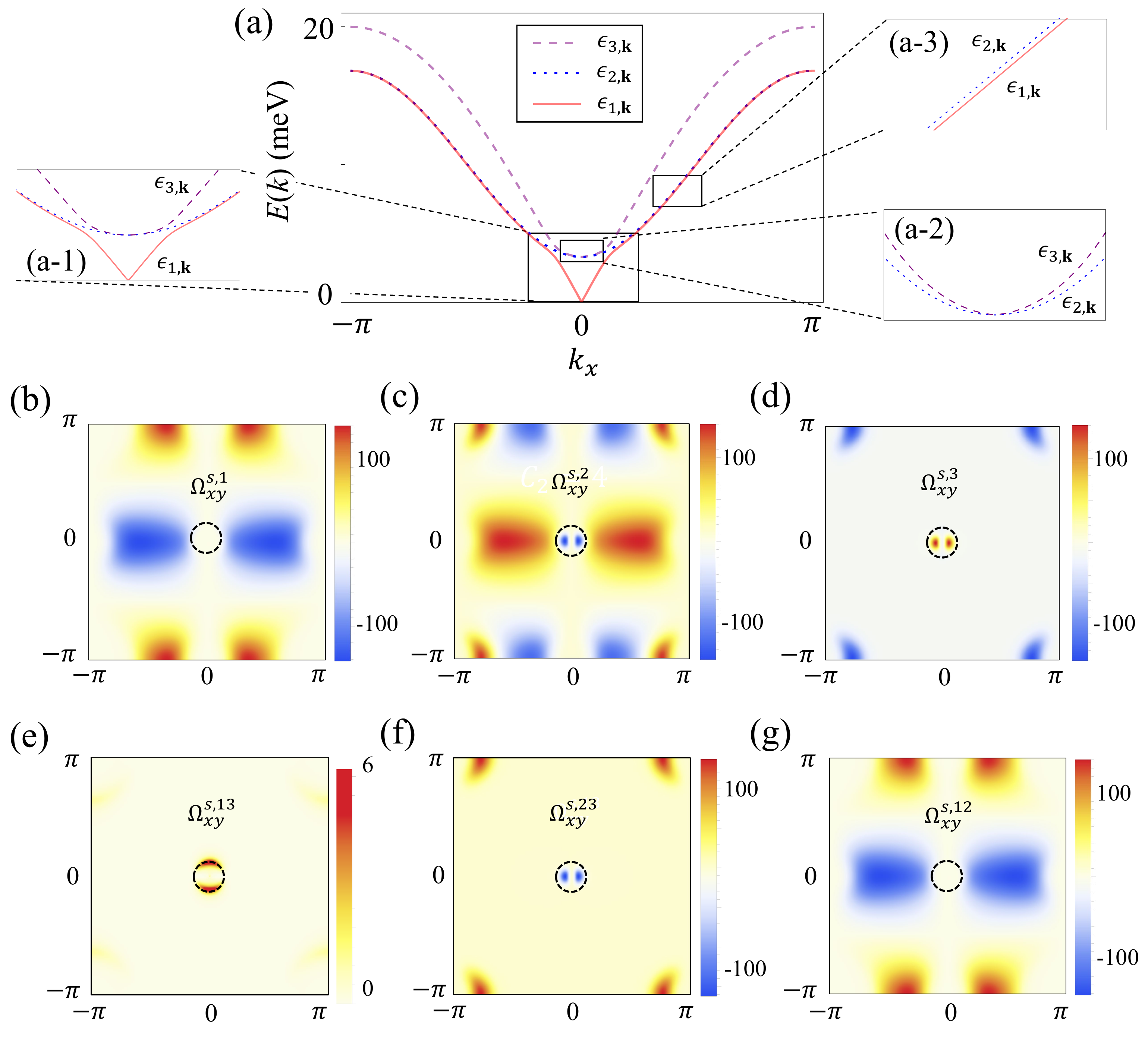}
\caption{The bulk band structure and spin Berry curvature in a layered AFM in the absence of an external magnetic field.
(a) The band structure of the three-band Hamiltonian. The square boxes show the band gaps where the spin Berry curvature can arise (zoomed in images).
Spin Berry curvatures of (b) lower (solid line), (c) middle (dotted line), and (d) upper (dashed line) bands.
Projected spin Berry curvatures from the interband transition between (e) lower and upper $(1\leftrightarrow 3)$, (f) middle and upper $(2\leftrightarrow 3)$, (g) lower and middle  $(1\leftrightarrow 2)$ bands. The dashed circles represent the anti-crossing points between a magnon and phonon bands.} \label{fig:2}
\end{figure}

\begin{equation}
\begin{aligned}\label{SBCII}
\Omega_{xy,\rm II}^{{s,n}}(\v k) &= \sum_{m\neq n}\left\{ \sum_{l\neq n} \frac{- \hbar^2{\rm Im} \left[ S_{z,\v k}^{nl} \langle l_{\v k}|v_x |m_\v k\rangle\langle m_{\v k}| v_y |n_{\v k}\rangle\right]}{(\epsilon_{n,\v k} - \epsilon_{m,\v k})^2}
-(n \leftrightarrow m)\right\} \equiv \sum_{m\neq n} \Omega_{xy,\rm II}^{{s,nm}}(\v k),
\end{aligned}
\end{equation}
\end{widetext}
where $S_{z,\v k}^{nl} = \langle n_{\v k}| S_z |l_{\v k}\rangle$ is the off-diagonal spin expectation.
On the contrary to the type-I spin Berry curvature, this term is not related to the topologically protected charge (Chern number).
In other words, the momentum integration of the type-II spin Berry curvature is not restricted to a quantized integer.
From the denominator of Eq.~\eqref{SBCII}, we read that the spin Berry curvature increases as the band gap decreases.
By using the projected spin Berry curvature $\Omega_{xy}^{{s,nm}}(\v k)$ in Eq.~\eqref{SBCII}, we provide a more detailed understanding of the spin Berry curvature.
As mentioned, the magnon-phonon interaction hybridizes the magnon and phonon bands and lifts the band-degeneracy.
A direct result of this is the band-anticrossing between a magnonic band and a phononic band (anti-crossing circle) which occurs at $\epsilon_{p,\v k} = \epsilon_{m, \v k}$ [dashed circles in Fig.~\ref{fig:2}(b)-(g)].
From the interband transition between the two bands ($\epsilon_{1, \v k}$ and $\epsilon_{3, \v k}$) the projected spin Berry curvature $\Omega_{xy}^{{s,13}}(\v k)$ becomes finite.
Because this effect is maximized at the anticrossing circle, the projected spin Berry curvature $\Omega_{xy}^{{s,13}}(\v k)$ is localized at these points [Fig.~\ref{fig:2}(e)].
More interesting point is that the spin Berry curvature can be finite apart from the anticrossing circle.
The magnon-phonon interaction generates slight band gaps between two magnonic bands which are degenerate without it.
As a result, the magnon-like bands possess a small phonon character over the whole Brillouin zone
causing the slight band gaps apart from the anticrossing circle [Fig.~\ref{fig:2}(a-2) and (a-3)].
Because these band gaps are much smaller than the anticrossing gap, the resultant projected spin Berry curvatures are large compared to that from the anticrossing circle [Fig.~\ref{fig:2}(f) and (g)].
This is one of our central results: The magnon-polaron band in the layered AFM can show the large spin Berry curvature and the ensuing topological spin transport even in the presence of the time-reversal symmetry.

\section{Thermal transport}\label{sec3}

\begin{figure}[t]
\includegraphics[width=70mm]{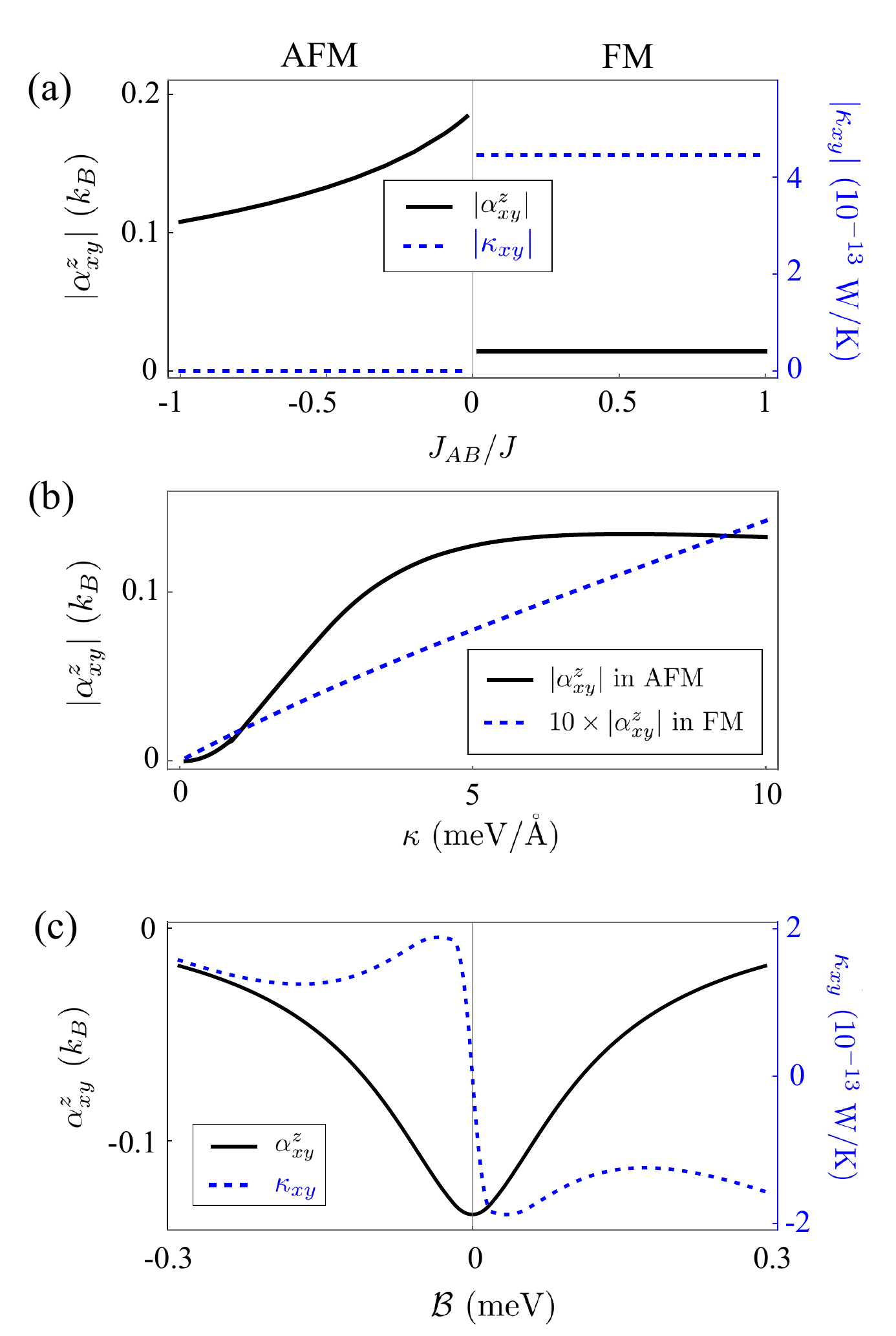}
\caption{Thermal response functions of the layered magnet.
(a) Interlayer exchange dependence of the spin Nernst and thermal Hall conductivities.
(b) Magnetoelastic interaction dependence of the spin Nernst conductivity.
(c) Magnetic field dependence of the spin Nernst and thermal Hall conductivities in the layered AFM.
For all figures, $\alpha = 0.001$ is used.
For (a) and (b), ${\cal B} = 0$ is used.
For (a) and (c), $\kappa = 10$ ${\rm meV}/{\rm \mathring{A}}$ is used.
For (b), $J_{AB} = -0.5 J$ and $J_{AB} = 0.5 J$ are used for AFM and FM, respectively.} \label{fig:3}
\end{figure}

Here we discuss the topological transverse transport of particle and spin angular momenta of the layered magnet under a temperature gradient.
The linear response equations for the thermal Hall and spin Nernst effects are given by~\cite{Matsumoto2011a, Matsumoto2011b, Li2020, Park2020}
\begin{align}
j^Q_y = -\kappa_{xy} \partial_x T,\qquad j^{S_z}_y = -\alpha^{z}_{xy} \partial_x T,
\end{align}
where
\begin{equation}
\begin{aligned}
&\kappa_{xy} = -\frac{k_B^2 T}{\hbar V} \sum_{n,\v k} c_2(\rho_n) \Omega^{n}_{xy}(\v k,\tau),\\
&\alpha^{z}_{xy} = \frac{k_B\hbar}{V} \sum_{n,\v k} c_1(\rho_n)\Omega_{xy}^{{s,n}}(\v k,\tau),
\end{aligned}
\end{equation}
are the corresponding conductivities, and
\begin{align}
\Omega^{n}_{xy}(\v k,\tau) = \sum_{m\neq n} \frac{{\rm Im} \left[-2\hbar^2 \langle n_\v k| v_x |m_\v k\rangle
\langle m_{\v k}| v_y |n_{\v k}\rangle\right]}{(\epsilon_{n,\v k} - \epsilon_{m,\v k})^2  + \left(\frac{\hbar}{\tau}\right)^2}.
\end{align}
Here, $c_1(\rho) = (1+ \rho){\rm log}(1+\rho) - \rho {\rm log} \rho$,
$c_2(\rho) = (1+\rho) \left[{\rm log} ({1+\rho}/{\rho})\right]^2 - ({\rm log} \rho)^2 - 2{\rm Li}_2(-\rho)$ where
$\rho_n = \left(e^{{\epsilon_n}/k_B T }-1\right)^{-1}$ is the Bose-Einstein distribution
function with a zero chemical potential, $k_B$ is the Boltzmann constant, $T$ is the temperature, and ${\rm Li}_2(\rho)$ is a polylogarithm function.

In Fig.~\ref{fig:3} we show the spin Nernst and thermal Hall conductivities of the layered FM and AFM with a finite relaxation time
(see Appendix~\ref{secA2} for detailed calculation in the layered FM).
For the relaxation time, we assume that the scattering is dominated by the Gilbert damping ${\tau_n} = 2\alpha \omega_{n,\v k}$,
which is valid for clean magnetic systems~\cite{Kovalev2012, Jiang2013, footnote1}.
Fig.~\ref{fig:3}(a) shows the spin Nernst and thermal Hall conductivities by changing the interlayer exchange which can be
controlled by applying the gate voltages~\cite{Jiang2018a,Jiang2018b,Huang2018,Morell2019}.
In the layered antiferromagnet, the thermal Hall conductivity vanishes due to the time-reversal symmetry but the spin Nernst conductivity is finite.
There are sudden changes of the spin Nernst and thermal Hall conductivities at the phase transition point between FM and AFM phases [Fig.~\ref{fig:3}(a)].
Interestingly, the spin Nernst conductivity in the AFM phase is about ten times larger than that in the FM phase [Fig.~\ref{fig:3}(b)].
The sudden increase (or decrease) of the spin Nernst conductivity at the phase transition point can be observed by measuring the voltage induced by the inverse spin Hall effect in the metallic leads~\cite{HZhang2022}.
The maximum value of the spin Nernst conductivity in the layered AFM is about $0.17$ $k_B$ at $T=40$ K.
We note that this value is quite large: $\alpha^z_{xy}$ in the AFM due to the magnon-phonon interaction is
two order of magnitude larger than $\alpha^z_{xy}$ obtained in previous studies~\cite{Cheng2016,Li2020}.
The giant spin Nernst conductivity is obtained for sufficiently long relaxation time because
the level broadening suppresses the energy splitting from the magnetoelastic interaction (see the Appendix~\ref{secA3} for the level broadening effect of the spin Nernst conductivity).
In Fig.~\ref{fig:3}(c), we also show the magnetic field dependence of the spin Nernst conductivity and thermal Hall conductivity in the layered AFM.
As expected, the spin Nernst conductivity is even in $\cal B$ $[\alpha^z_{xy} ({\cal B}) = \alpha^z_{xy} (-{\cal B})]$ while thermal Hall conductivity is odd in
$\cal B$ $[\kappa_{xy} ({\cal B}) = -\kappa_{xy} (-{\cal B})]$ (see the Appendix~\ref{secA4}).

\section{Discussion}\label{sec4}

In this paper, we study topological spin transport of the magnon-polaron bands in the bilayer system with two FM monolayers.
The magnetoelastic interaction opens the band gaps between different excitation bands and generates the spin Berry curvature even if the time-reversal symmetry is preserved.
Previous studies~\cite{Go2019, SZhang2020} report that the magnetoelastic interaction opens band-anticrossing gaps between a magnonic and a phononic bands.
In our theory, the dominant contribution to the spin Berry curvature does not originate from the band-anticrossings between a magnonic and a phononic bands.
Due to the magnetoelastic interaction, the originally degenerate two magnonic bands ($\alpha$- and $\beta$-magnons) hybridizes with the phononic band.
As a result, there are slight band gaps between two magnon-like bands distant from the anticrossing circle.
Because these band gaps are much smaller than the anticrossing gap, they give the dominant contribution to the spin Berry curvature.
For an experimental probe of our predictions, we compute the spin Nernst and thermal Hall conductivities in our model.
We find that the spin Nernst conductivity increases significantly when the interlayer exchange becomes antiferromagnetic.
We note that the large spin Nernst conductivity from the small band gap is analogous to the additional intrinsic spin-orbit torque in the antiferromagnet with a small noncolinearity~\cite{Cheon2018}.
The phase tunability of the magnetic bilayer by the external electric field~\cite{Jiang2018a,Jiang2018b,Huang2018,Morell2019} possibly provide an efficient control scheme of the energy and angular momentum transfer for future spintronic devices.
We end this paper by noting that our theory is applicable for general AFM materials.
As mentioned, the effective Hamiltonian of the antiferromagnetically coupled bilayer is equivalent to that of the monolayer AFM~\cite{SZhang2020} in low energy limit.
Also, our theoretical model is based on the simple square lattice magnet and does not require specific lattice structure.
These suggest that the ubiquity of topological spin transports in conventional AFMs with reduced dimensions.

\begin{acknowledgments}
We thank Kyoung-Whan Kim and Jung Hyun Oh for the useful discussions. G.G. acknowledges a support by the National Research Foundation of Korea (NRF-2022R1C1C2006578).
S.K.K. was supported by Brain Pool Plus Program through the National Research Foundation of Korea funded by the Ministry of Science and ICT (NRF-2020H1D3A2A03099291) and National Research Foundation of Korea funded by the Korea Government via the SRC Center for Quantum Coherence in Condensed Matter (NRF-2016R1A5A1008184).
\end{acknowledgments}

\appendix
\counterwithin{figure}{section}

\section{Derivation of the effective Hamiltonian in the bilayer antiferromagnet}\label{secA1}

\subsection{Magnon part}

We divide the magnetic Hamiltonian into the intralayer and interlayer parts $H_{m} = H^0_m + H^1_m$, where
\begin{align}
H^0_m &= -J \sum_{l \in \{A,B\}}\sum_{\langle i,j\rangle} \v S^l_i\cdot \v S^l_j - \frac{K_z}{2} \sum_{i,l} (S^l_{i,z})^2 - {\cal B} \sum_{i,l} S^l_{i,z},\nn
H^1_m &= - J_{AB} \sum_i \v S^A_i\cdot \v S^B_i.
\end{align}
By using the Holstein-Primakoff transformation
\begin{equation}
\begin{aligned}
&S^+_A = \sqrt{2S} a_A,\quad S^-_A = \sqrt{2S} a^\dag_A.\quad S^z_A = S- a^\dag_A a_A,\nn
&S^+_B = \sqrt{2S} a^\dag_B,\quad S^-_B = \sqrt{2S} a_B.\quad S^z_B = a^\dag_B a_B - S,
\end{aligned}
\end{equation}
and the Fourier transformation $a_i = \sum_{\v k} e^{i \v k\cdot R_i} a_{\v k}/N$, we have
\begin{equation}
\begin{aligned}
H^0_m =&\sum_{\v k} \left[\epsilon^A_{m,\v k} a^\dag_{A, \v k} a_{A, \v k} + \epsilon^B_{m,\v k} a^\dag_{B, \v k} a_{B, \v k}\right],
\end{aligned}
\end{equation}
where $\epsilon^{A/B}_{m,\v k} = 2JS\left(2-\cos{k_x} - \cos{k_y}\right) + K_zS \pm {\cal B}$ and
\begin{widetext}
\begin{equation}
\begin{aligned}
H^1_m &= |J_{\rm AB}|\sum_i {\v S}^A_i \cdot {\v S}^B_i = |J_{\rm AB}|\sum_i \left[\frac{1}{2} \left(S^+_{A,i} S^-_{B,i} + S^-_{A,i} S^+_{B,i}\right) + S^z_{A,i} S^z_{B,i}\right]\nn
& \approx |J_{\rm AB}| S\sum_\v k \left(a_{A,\v k}a_{B,-\v k} + a^\dag_{A,\v k}a^\dag_{B,-\v k} + a^\dag_{A,\v k}a_{A,\v k} + a^\dag_{B,\v k}a_{B,\v k} \right).
\end{aligned}
\end{equation}
By using the Bogoliubov transformation,
\begin{align}
a_{A, \v k} = u_{\v k} \alpha_{\v k} + v_{\v k} \beta^\dag_{-\v k},\quad a_{B, \v k} = u_{\v k} \beta_{\v k} + v_{\v k} \alpha^\dag_{-\v k},
\end{align}
we have
\begin{equation}
\begin{aligned}
H_m
=&\sum_{\v k}\Bigg\{\left[ \left(\epsilon_{m,\v k}^A + |J_{\rm AB}| S\right) u_{\v k}^2 + \left(\epsilon_{m,\v k}^B
+ |J_{\rm AB}| S\right) v_{\v k}^2 + 2 |J_{\rm AB}| S u_{\v k} v_{\v k} \right] \alpha^\dag_{\v k} \alpha_{\v k}\\
&\hspace{10mm}+ \left[ \left(\epsilon_{m,\v k}^A + |J_{\rm AB}| S\right) v_{\v k}^2 + \left(\epsilon_{m,\v k}^B+ |J_{\rm AB}| S\right) u_{\v k}^2 + 2 |J_{\rm AB}| S u_{\v k} v_{\v k} \right] \beta^\dag_{\v k} \beta_{\v k}\\
&\hspace{10mm}+\left[ \left(\epsilon_{m,\v k}^A + \epsilon_{m,\v k}^B + 2|J_{\rm AB}| S\right) u_{\v k} v_{\v k} +  |J_{\rm AB}| S \left(u^2_{\v k} + v^2_{\v k}\right) \right]
(\alpha_{\v k} \beta_{-\v k}+ \alpha^\dag_{\v k} \beta^\dag_{-\v k})\Bigg\}.
\end{aligned}
\end{equation}
\end{widetext}
In order to eliminate the $\alpha_{\v k} \beta_{-\v k}+ \alpha^\dag_{\v k} \beta^\dag_{-\v k}$ terms, we set
\begin{align}
&u_{\v k} = \cosh\theta_{\v k}, \quad v_{\v k} = \sinh\theta_{\v k},\nn
&\tanh 2\theta_{\v k} = -\frac{2 |J_{\rm AB}| S}{\epsilon_{m,\v k}^A + \epsilon_{m,\v k}^B + 2 |J_{\rm AB}| S}.\nonumber
\end{align}
Note that both $u_{\v k}$ and $v_{\v k}$ are real and $u_{-\v k} = u_{\v k}$ and $v_{-\v k} = v_{\v k}$.
Then, we have
\begin{align}\label{Hmagf}
H_m =\sum_{\v k} \left (\epsilon^\alpha_{m,\v k} \alpha^\dag_{\v k} \alpha_{\v k} + \epsilon^\beta_{m,\v k} \beta^\dag_{\v k} \beta_{\v k}\right),
\end{align}
where $\epsilon^{\alpha/\beta}_{m,\v k} = \sqrt{(\epsilon^0_{m,\v k})^2 + 2 \epsilon^0_{m,\v k}|J_{\rm AB}| S} \pm {\cal B}$
and $\epsilon^0_{m,\v k} = 2JS\left(2-\cos{k_x} - \cos{k_y}\right) + K_zS$.

\subsection{Phonon part}

The momentum space representation of the phonon Hamiltonian is
\begin{align}
H_{ph} &= \sum_{\v k} \left[\frac{p^z_{-\v k} p^z_{\v k}}{2M} + \frac{1}{2} u^z_{-\v k} \Phi(\v k) u^z_{\v k}\right],
\end{align}
where
\begin{align}
\Phi(\v k) = &\left(M \omega_{p,\v k}^2  + M \omega_z^2\right) I_{2\times 2} + M \omega_z^2 \left(
                                                                                                \begin{array}{cc}
                                                                                                  0 & 1 \\
                                                                                                  1 & 0 \\
                                                                                                \end{array}
                                                                                              \right).
\end{align}
The spring constant matrix $\Phi(\v k)$ can be diagonalized by the similarity transformation ${\cal S}^{-1} \Phi(\v k) {\cal S}$, where
\begin{align}
{\cal S} = \frac{1}{\sqrt2}\left(
             \begin{array}{cc}
               1 & 1 \\
               1 & -1 \\
             \end{array}
           \right).
\end{align}
By introducing the canonical quantization of the lattice vibration and conjugate momentum in the diagonalized basis,
\begin{align}
u^z_{i,\v k} &= \sqrt{\frac{\hbar}{M \omega_{p,i}(\v k)}} \left(\frac{b_{i,\v k} + b^\dag_{i,-\v k}}{\sqrt2}\right),\\
p^z_{i,\v k} &= \sqrt{\hbar M \omega_{p,i}(\v k)} \left(\frac{b_{i,-\v k} - b^\dag_{i,\v k}}{\sqrt2i}\right),
\end{align}
we have
\begin{align}\label{Hphf}
H_{ph} &= \sum_{\v k} \left[\hbar \omega_{p,1} (\v k) b^\dag_{1,\v k} b_{1,\v k} + \hbar \omega_{p,2} (\v k) b^\dag_{2,\v k} b_{2,\v k}\right],
\end{align}
where $\omega_{p,1} (\v k) = \sqrt{\omega_{p,\v k}^2 + 2 \omega_z^2}$ and $\omega_{p,2} (\v k) = \omega_{p, \v k}$ are frequencies of the anti-bonding and bonding phonons, respectively.

\begin{widetext}
\subsection{Magnon-phonon interaction}

The magnon-phonon interaction in the bilayer system is
\begin{equation}
\begin{aligned}
H_{mp} =& \kappa \sum_i \Bigg\{ \left[ S^A_{i,x}\left(u^z_{A,i} - u^z_{A,i+{a \hat {\v x}}}\right) + S^A_{i,y}\left(u^z_{A,i} - u^z_{A,i+a \hat {\v y}}\right)
- S^A_{i,x}\left(u^z_{A,i} - u^z_{A,i-{a \hat {\v x}}}\right) - S^A_{i,y}\left(u^z_{A,i} - u^z_{A,i-a \hat {\v y}}\right)\right]\nn
&\hspace{12mm}- \left[ S^B_{i,x}\left(u^z_{B,i} - u^z_{B,i+{a \hat {\v x}}}\right) + S^B_{i,y}\left(u^z_{B,i} - u^z_{B,i+a \hat {\v y}}\right)
- S^B_{i,x}\left(u^z_{B,i} - u^z_{B,i-{a \hat {\v x}}}\right) - S^B_{i,y}\left(u^z_{B,i} - u^z_{B,i-a \hat {\v y}}\right)\right]\Bigg\}\\
=&  2\kappa \sqrt{S} \sum_{\v k}\left[ i \sin k_x u^z_{A,-\v k} \left(\frac{ a_{A,\v k}+ a^\dag_{A,-\v k} }{\sqrt2}\right)
+ i \sin k_y u^z_{A,-\v k} \left(\frac{ a_{A,\v k}- a^\dag_{A,-\v k}}{\sqrt2 i}  \right) \right]\\
&-2\kappa \sqrt{S} \sum_{\v k}\left[ i \sin k_x u^z_{B,-\v k} \left(\frac{ a_{B,\v k}+ a^\dag_{B,-\v k} }{\sqrt2}\right)
- i \sin k_y u^z_{B,-\v k} \left(\frac{ a_{B,\v k}- a^\dag_{B,-\v k}}{\sqrt2 i}  \right) \right].
\end{aligned}
\end{equation}
By acting the similarity transformation which diagonalizes the spring constant matrix, we have
\begin{equation}
\begin{aligned}
{H}_{mp}
= &  2\kappa \sqrt{S} \sum_{\v k}\left[ i \sin k_x \left(\frac{u^z_{1,-\v k} + u^z_{2,-\v k}}{\sqrt2}\right)\left(\frac{ a_{A,\v k}+ a^\dag_{A,-\v k} }{\sqrt2}\right)
+ i \sin k_y \left(\frac{u^z_{1,-\v k} + u^z_{2,-\v k}}{\sqrt2}\right) \left(\frac{ a_{A,\v k}- a^\dag_{A,-\v k}}{\sqrt2 i}  \right) \right]\nn
&-2\kappa \sqrt{S} \sum_{\v k}\left[ i \sin k_x \left(\frac{u^z_{1,-\v k} - u^z_{2,-\v k}}{\sqrt2}\right) \left(\frac{ a_{B,\v k}+ a^\dag_{B,-\v k} }{\sqrt2}\right)
- i \sin k_y \left(\frac{u^z_{1,-\v k} - u^z_{2,-\v k}}{\sqrt2}\right) \left(\frac{ a_{B,\v k}- a^\dag_{B,-\v k}}{\sqrt2 i}  \right) \right]
\end{aligned}
\end{equation}
By using $a_{A, \v k} = u_{\v k} \alpha_{\v k} + v_{\v k} \beta^\dag_{-\v k}$, $a_{B, \v k} = u_{\v k} \beta_{\v k} + v_{\v k} \alpha^\dag_{-\v k}$ and $\displaystyle{u^z_{i,-\v k} = \sqrt{\frac{\hbar}{M \omega_{p,i}(\v k)}} \left(\frac{b_{i,-\v k} + b^\dag_{i,\v k}}{\sqrt2}\right)}$,
we have
\begin{align}
H = H_{mp}^1 + H_{mp}^2,
\end{align}
where
\begin{align}
{H}^1_{mp}&=\frac{{\kappa}}{2} \sum_{\v k} \sqrt{\frac{2\hbar S}{M \omega_{p,1}(\v k)}} (u_{\v k}- v_{\v k}) \psi^\dag_{1,\v k}
                \left(
                  \begin{array}{cccccc}
                    0 & 0 & 0 & 0 & i \sin k_x  & 0 \\
                    0 & 0 & 0 & 0 & i \sin k_y  & 0 \\
                    0 & 0 & 0 & 0 & -i \sin k_x  & 0 \\
                    0 & 0 & 0 & 0 & i \sin k_y  & 0 \\
                    -i \sin k_x  & -i \sin k_y  & i \sin k_x  & -i \sin k_y & 0 & 0 \\
                    0 & 0 & 0 & 0 & 0 & 0 \\
                  \end{array}
                \right)\psi_{1,\v k},\nn
{H}^2_{mp}&=\frac{{\kappa}}{2} \sum_{\v k} \sqrt{\frac{2\hbar S}{M \omega_{p,2}(\v k)}} (u_{\v k}+ v_{\v k}) \psi^\dag_{2,\v k}
                \left(
                \begin{array}{cccccc}
                    0 & 0 & 0 & 0 & i \sin k_x  & 0 \\
                    0 & 0 & 0 & 0 & i \sin k_y  & 0 \\
                    0 & 0 & 0 & 0 & i \sin k_x  & 0 \\
                    0 & 0 & 0 & 0 & -i \sin k_y  & 0 \\
                    -i \sin k_x  & -i \sin k_y  & -i \sin k_x  & i \sin k_y & 0 & 0 \\
                    0 & 0 & 0 & 0 & 0 & 0 \\
                  \end{array}
                \right)\psi_{2,\v k},
\end{align}
and
\begin{align}
&\psi_{i,\v k} = \left(\frac{\alpha_{\v k} + \alpha^\dag_{-\v k}}{\sqrt2}, \frac{\alpha_{\v k} - \alpha^\dag_{-\v k}}{\sqrt2 i}, \frac{\beta_{\v k} + \beta^\dag_{-\v k}}{\sqrt2}, \frac{\beta_{\v k} - \beta^\dag_{-\v k}}{\sqrt2 i},
\frac{b_{i,-\v k} + b^\dag_{i,\v k}}{\sqrt2}, \frac{b_{i,-\v k} - b^\dag_{i,\v k}}{{\sqrt2} i}\right)^T,
\qquad \psi_{i,\v k}^\dag = \psi_{i,-\v k}.
\end{align}
Therefore, the topology of the magnon-polaron in bilayer antiferromagnet is described by the $8\times8$ matrix.
Neglecting the high energy phonon state (anti-bonding phonon), and collecting the particle number conserving terms, we have the effective three-band Hamiltonian of the magnetoelastic interaction
\begin{align}\label{Hmpf}
{H}_{mp}& = {\kappa} \sum_{\v k} \sqrt{\frac{\hbar S}{2M \omega_{p}(\v k)}} (u_{\v k}+ v_{\v k})
\psi^\dag_{\v k} \left(
  \begin{array}{ccc}
    0 & 0 & i \sin k_x + \sin k_y \\
    0 & 0 & i \sin k_x - \sin k_y \\
    -i \sin k_x + \sin k_y & -i \sin k_x  - \sin k_y & 0 \\
  \end{array}
\right)\psi_{\v k},
\end{align}
where $\psi_{\v k} = (\alpha_{\v k}, \beta_{\v k}, b_{\v k})$.
After some algebra, we have
\begin{align}
u_{\v k} \pm v_{\v k} &= \sqrt{\cosh 2\theta_{\v k} \pm \sinh 2\theta_{\v k} } = \left(\frac{1+\delta \pm \delta}{\sqrt{1+2\delta}} \right)^{1/2}.
\end{align}
where $\delta = {2J_{\rm AB} S}/{\epsilon^0_{m,\v k}}$.

\subsection{Effective Hamiltonian}

By collecting Eqs.~\eqref{Hmagf}, ~\eqref{Hphf}, and ~\eqref{Hmpf}, we obtain the effective three-band Hamiltonian of the layered antiferromagnet
\begin{align}
H = H_{m} + H_{p} + H_{mp},
\end{align}
where
\begin{align}\label{Htot2}
{H}& = \sum_{\v k}
\psi^\dag_{\v k}
\left(
  \begin{array}{ccc}
    \epsilon^\alpha_{m, \v k} (\v k) & 0 & {\bar\kappa (\v k)}(i \sin k_x + \sin k_y) \\
    0 & \epsilon^\beta_{m, \v k} & {\bar\kappa (\v k)}(i \sin k_x - \sin k_y) \\
    {\bar\kappa (\v k)}(-i \sin k_x + \sin k_y) & {\bar\kappa_2 (\v k)}(-i \sin k_x - \sin k_y )& \epsilon_{p,\v k} \\
  \end{array}
\right)\psi_{\v k},
\end{align}
where
\begin{align}
{\bar \kappa} = \kappa \sqrt{\frac{\hbar S\sqrt{(1+2\delta)}}{2M {\omega}_{p,\v k}}}.
\end{align}
In Fig.~\ref{fig:S1}, we show that the band structure of the effective three-band Hamiltonian and that of the full $8\times8$ Hamiltonian.
The highest state of the full $8\times8$ Hamiltonian is the anti-bonding phonon mode which is neglected in the effective Hamiltonian.
The band structure in both models are almost the same for small $\kappa$.
\begin{figure}[h]
\includegraphics[width=110mm]{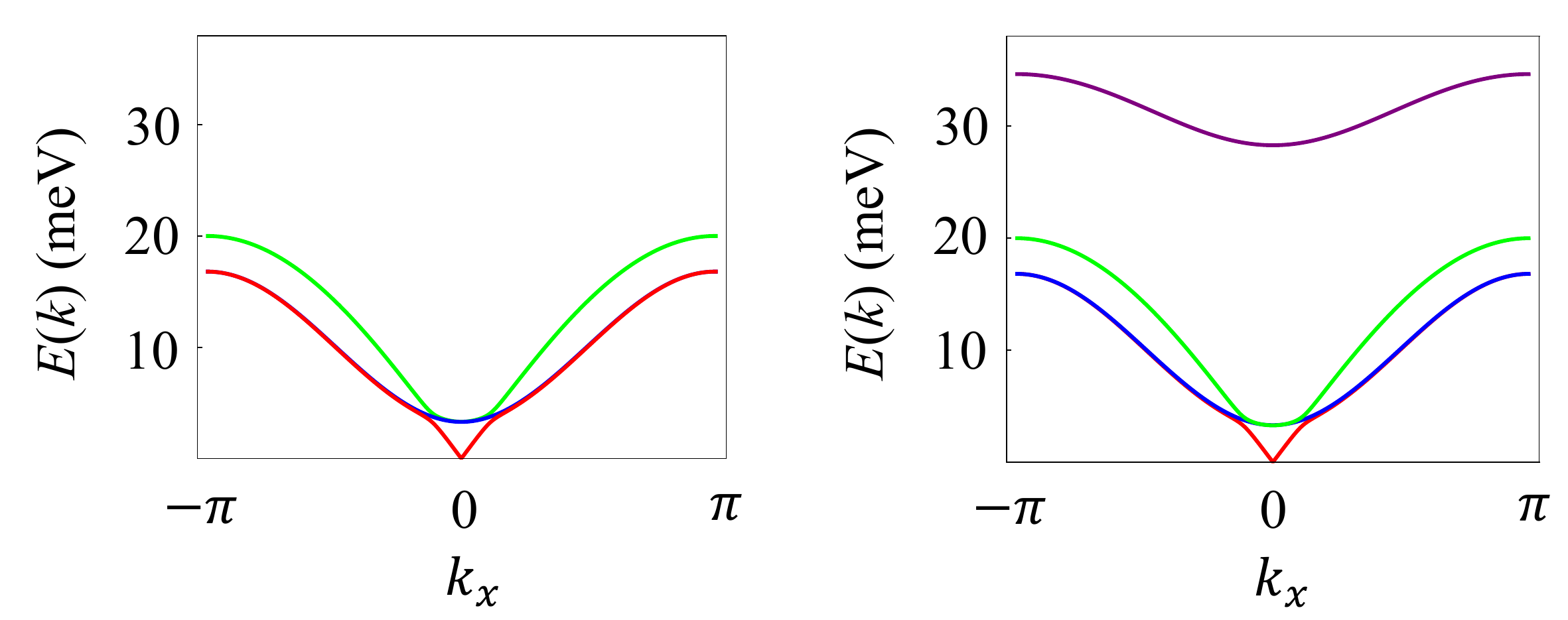}
\caption{The bulk band structure of the effective three-band Hamiltonian (left) and that of the full $8\times8$ Hamiltonian (right).
Material parameters are $J= 2.2$ meV, $K_z = 1.36$ meV, $S = 3/2$, $M c^2 = 5\times 10^{10}$ eV, $\hbar \omega_0 = 10$ meV, $\kappa = 5$ ${\rm meV}/{\rm \mathring{A}}$,
$\omega_z = 2 \omega_0$ and $J_{\rm AB} = -J/2$.}\label{fig:S1}
\end{figure}

\section{Derivation of the effective Hamiltonian in the bilayer ferromagnet}\label{secA2}

When the interlayer Heisenberg exchange is ferromagnetic, we have
\begin{align}
H_{m} &= \sum_{\v k} \left[(\epsilon^0_{m,\v k} + |J_{\rm AB}| S)(a^\dag_{A,\v k} a_{A,\v k} + a^\dag_{B,\v k} a_{B,\v k})
- J_{AB} S (a^\dag_{B,\v k} a_{A,\v k} + a^\dag_{A,\v k} a_{B,\v k})\right],\nn
&= \sum_{\v k} \psi^\dag_m \left[(\epsilon^0_{m,\v k} + |J_{\rm AB}| S)I_{2\times 2} - |J_{\rm AB}| S \left(
                                                                                                \begin{array}{cc}
                                                                                                  0 & 1 \\
                                                                                                  1 & 0 \\
                                                                                                \end{array}
                                                                                              \right)\right] \psi_m,
\end{align}
where $\psi_m = (a_{A,\v k}, a_{B,\v k})$. We note the the phonon Hamiltonian is the same as that in the layered antiferromagnet.
The submatrices ${\cal H}_{\rm mag}$ can be diagonalized by the similarity transformations ${\cal S}^{-1} {\cal H}_{\rm mag} {\cal S}$.
In this new basis, we have
\begin{align}
H_{m} &= \sum_{\v k} \left[(\epsilon^0_{m,\v k} + 2|J_{\rm AB}| S)a^\dag_{1,\v k} a_{1,\v k} + \epsilon^0_{m,\v k} a^\dag_{2,\v k} a_{2,\v k}\right].
\end{align}
We note that the magnon-phonon interaction
\begin{align}
H_{mp} &= \sum_i \sum_{{\v e}_i} ({\v S}_i \cdot {\v e}_i)  (u^z_i - u^z_{i+ {\v e}_i})\nn
&=  2\kappa \sqrt{S} \sum_{\v k}\left[ i \sin k_x u_{-\v k} \left(\frac{ a_{\v k}+ a^\dag_{-\v k} }{\sqrt2}\right)
+ i \sin k_y u_{-\v k} \left(\frac{ a_{\v k}- a^\dag_{-\v k}}{\sqrt2 i}  \right) \right]
\end{align}
is invariant under the basis transformation, i.e. ${\cal H}^{mp}_{\v k} = {\cal S}^{-1} {\cal H}^{mp}_{\v k} {\cal S}$.
Then, we write the total Hamiltonian in the new basis
\begin{align}
{\cal H}_{\v k} = \left(
      \begin{array}{cccccccc}
        \epsilon^0_{m,\v k} + 2|J_{\rm AB}| S & 0 & M_1^1 & 0 & 0 & 0 & 0 & 0 \\
        0 & \epsilon^0_{m,\v k} + 2|J_{\rm AB}| S & M^1_2 & 0 & 0 & 0 & 0 & 0 \\
        (M_1^1)^\ast & (M_2^1)^\ast & \hbar \omega_{p,1} (\v k)   & 0 & 0 & 0 & 0 & 0 \\
        0 & 0 & 0 & \hbar \omega_{p,1} (\v k) & 0 & 0 & 0 & 0 \\
        0 & 0 & 0 & 0 & \epsilon^0_{m,\v k} & 0 & M_1^2 & 0 \\
        0 & 0 & 0 & 0 & 0 & \epsilon^0_{m,\v k}  & M_2^2 & 0 \\
        0 & 0 & 0 & 0 & (M_1^2)^\ast & (M_2^2)^\ast & \hbar \omega_{p,2} (\v k) & 0 \\
        0 & 0 & 0 & 0 & 0 & 0 & 0 & \hbar \omega_{p,2} (\v k)\\
      \end{array}
    \right),
\end{align}
where
\begin{align}
(M_1)^i = 2i \kappa \sqrt{\frac{\hbar S}{M \omega_{i, \v k}}} \sin k_x ,\qquad (M_2)^i = -2i \kappa  \sqrt{\frac{\hbar S}{M \omega_{i, \v k}}} \sin k_y\qquad (i = 1,2).
\end{align}
By neglecting the high-frequency anti-bonding phonon and particle-number-nonconserving terms, we have
\begin{align}\label{Hbi}
H =\sum_{\v k} \left(
                      \begin{array}{cccc}
                        a^\dag_{1,\v k} & a^\dag_{2,\v k} & b^\dag_{2,\v k} \\
                      \end{array}
                    \right) {\cal H_{\v k}}  \left(
                    \begin{array}{c}
                      a_{1,\v k}  \\
                      a_{2,\v k}  \\
                      b_{2,\v k} \\
                    \end{array}
                  \right),
\end{align}
where
\begin{align}
{\cal H_{\v k}} = \left(
                    \begin{array}{ccc}
                      \epsilon^0_{m,\v k} + 2|J_{\rm AB}| S  & 0 & 0 \\
                      0 & \epsilon^0_{m,\v k}  & \tilde \kappa (\sin k_y - i \sin k_x) \\
                      0 &  \tilde \kappa (\sin k_y + i \sin k_x) & \hbar \omega_{p,2}(\v k) \\
                    \end{array}
                  \right),
\end{align}
and $\tilde \kappa = \kappa\sqrt{\frac{\hbar S}{M\omega_{p,2}(\v k) }} $.

\section{Effect of the level broadening}\label{secA3}

Because the dominant contribution of the spin Berry curvature in the antiferromagnet comes from the small energy spacing between two magnon-like bands ($\alpha$- and $\beta$-magnons),
the spin Nernst conductivity is suppressed by the level broadening effect. To see this, we show the damping constant dependence of the spin Nernst conductivity in Fig.~\ref{fig:S5}.
\begin{figure}[h]
\includegraphics[width=75mm]{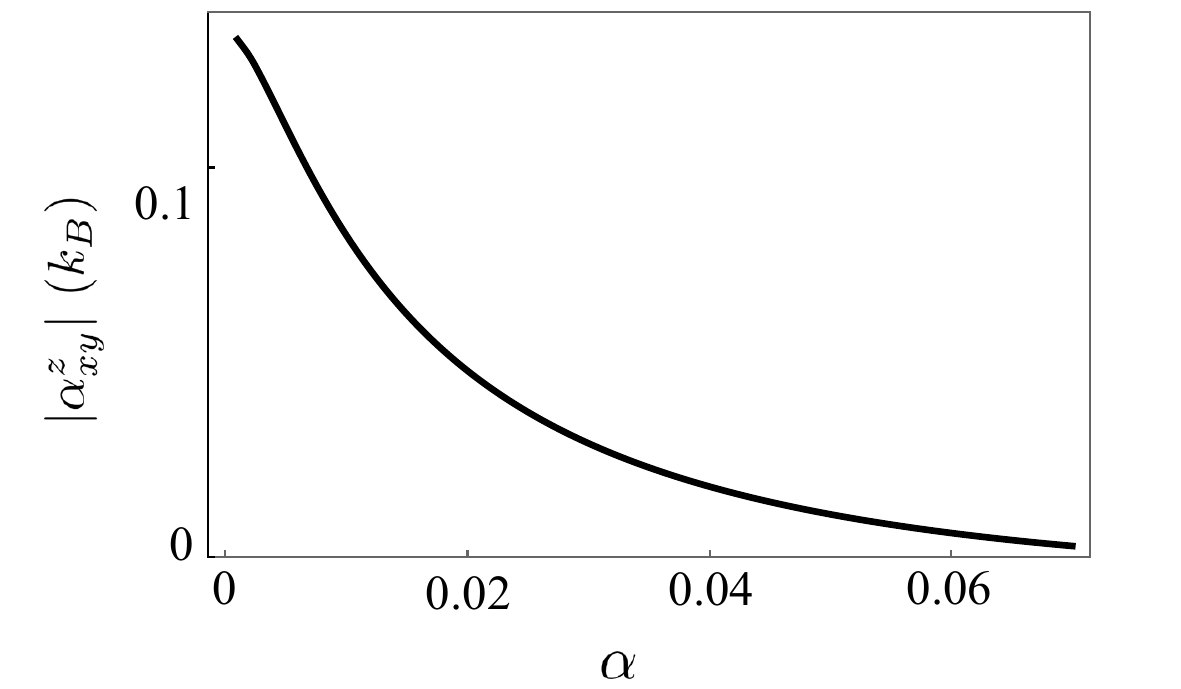}
\caption{Damping constant dependence of the spin Nernst conductivity without the external magnetic field. For the material parameters, we use $J= 2.2$ meV, $K_z = 1.36$ meV, $S = 3/2$, $M c^2 = 5\times 10^{10}$ eV, $\hbar \omega_0 = 10$ meV, $\kappa = 10$ ${\rm meV}/{\rm \mathring{A}}$, $\omega_z = 2 \omega_0$ and $J_{\rm AB} = -J/2$.}\label{fig:S5}
\end{figure}

\section{Effect of the external magnetic field}\label{secA4}

As discussed in Ref.~\cite{SZhang2020}, the nontrivial topology of the magnon-polaron (characterized by the Chern number) is induced by the external magnetic field.
Thus, the signs of the Berry curvatures are reversed when the sign of the magnetic field is reversed.
However, the spin Berry curvature is finite without the magnetic field and does not change under a sign reversal of the external magnetic field (see Fig.~\ref{fig:S2} and \ref{fig:S3}).

\begin{figure}[h]
\includegraphics[width=108mm]{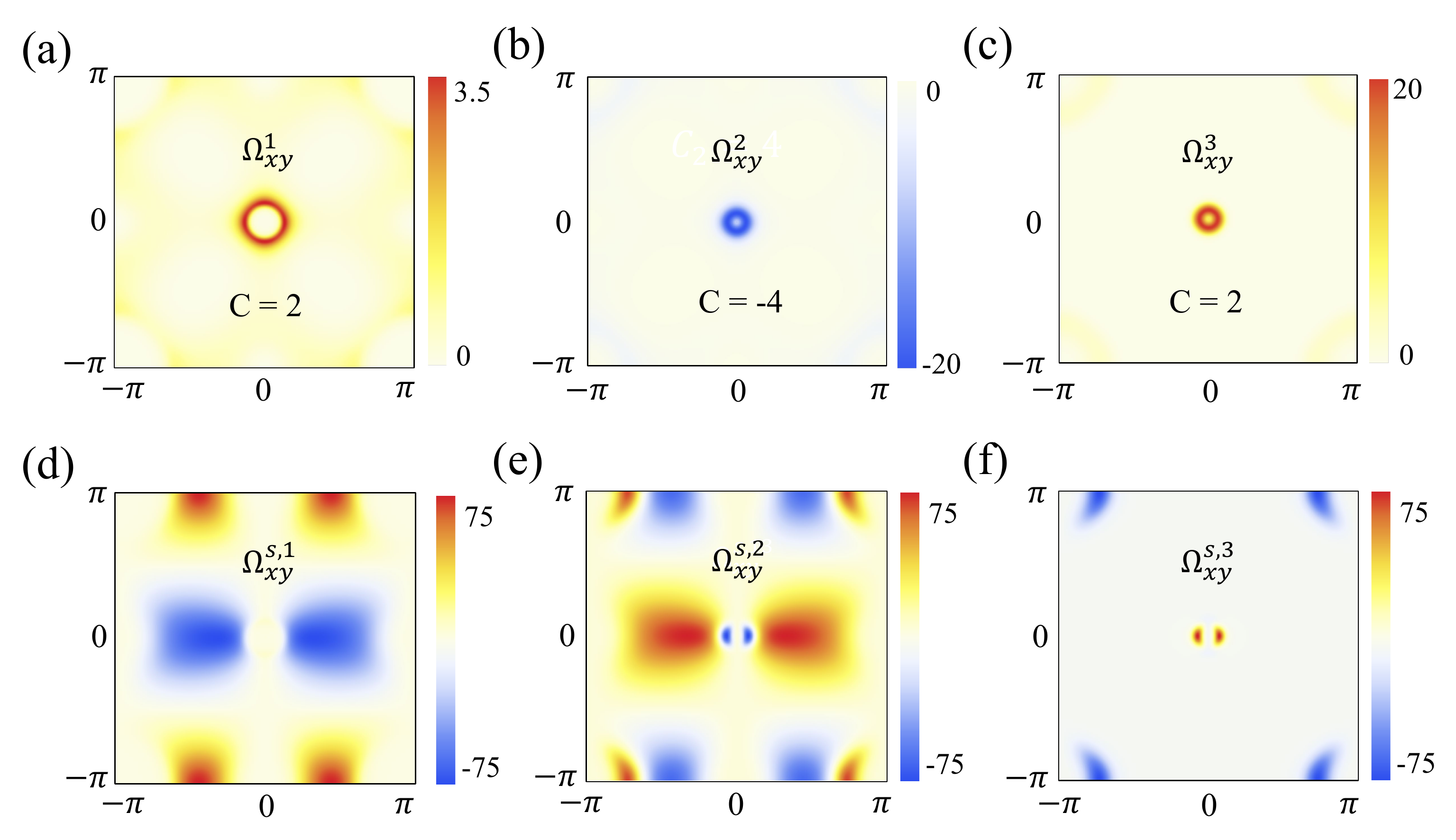}
\caption{Berry curvatures and spin Berry curvatures under the magnetic field ${\cal B} = 0.02$ meV. For the material parameters, we use $J= 2.2$ meV, $K_z = 1.36$ meV, $S = 3/2$, $M c^2 = 5\times 10^{10}$ eV, $\hbar \omega_0 = 10$ meV, $\kappa = 5$ ${\rm meV}/{\rm \mathring{A}}$, $\omega_z = 2 \omega_0$ and $J_{\rm AB} = -J/2$.}\label{fig:S2}
\end{figure}
\begin{figure}[h]
\includegraphics[width=108mm]{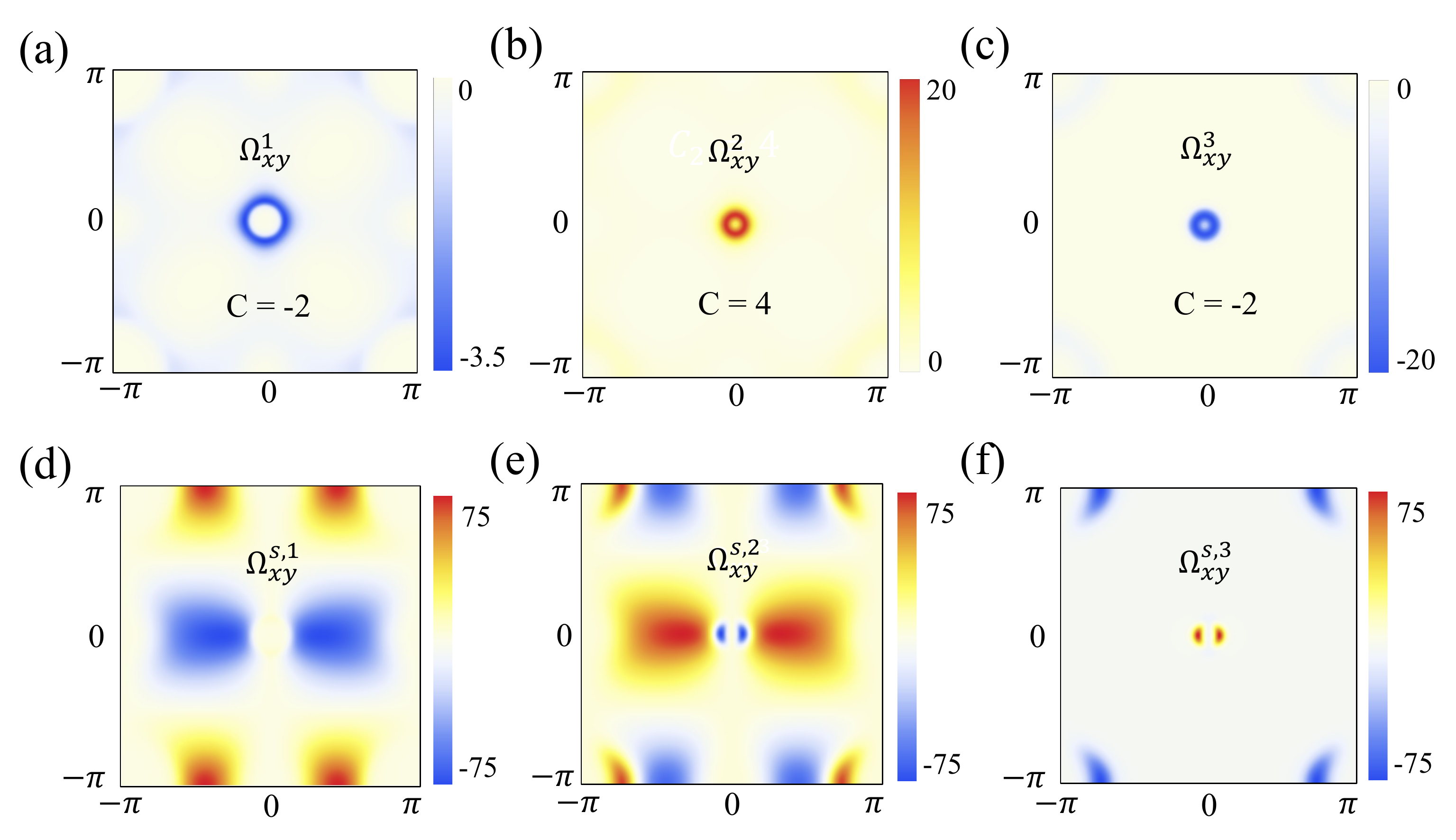}
\caption{Berry curvatures and spin Berry curvatures under the magnetic field ${\cal B} = -0.02$ meV. For the material parameters, we use $J= 2.2$ meV, $K_z = 1.36$ meV, $S = 3/2$, $M c^2 = 5\times 10^{10}$ eV, $\hbar \omega_0 = 10$ meV, $\kappa = 5$ ${\rm meV}/{\rm \mathring{A}}$, $\omega_z = 2 \omega_0$ and $J_{\rm AB} = -J/2$.}\label{fig:S3}
\end{figure}

\end{widetext}


\newpage


\begin{thebibliography}{99}

\bibitem{Chumak2015} A. V. Chumak, V. I. Vasyuchka, A. A. Serga, and B. Hillebrands, Nat. Phys. \textbf{11}, 453 (2015).
\bibitem{Maldovan2013} M. Maldovan, Nature \textbf{503}, 209 (2013).

\bibitem{Katsura2010} H. Katsura, N. Nagaosa, and P. A. Lee, Phys. Rev. Lett. \textbf{104}, 066403 (2010).
\bibitem{Onose2010} Y. Onose, T. Ideue, H. Katsura, Y. Shiomi, N. Nagaosa, and Y. Tokura, Science \textbf{329}, 297 (2010).
\bibitem{Matsumoto2011a} R. Matsumoto and S. Murakami, Phys. Rev. Lett. \textbf{106}, 197202 (2011).
\bibitem{Matsumoto2011b} R. Matsumoto and S. Murakami, Phys. Rev. B \textbf{84}, 184406 (2011).
\bibitem{Shindou2013a} R. Shindou, R. Matsumoto, S. Murakami, and J.-i. Ohe, Phys. Rev. B \textbf{87}, 174427 (2013).
\bibitem{Shindou2013b} R. Shindou, J.-i. Ohe, R. Matsumoto, S. Murakami, and E. Saitoh, Phys. Rev. B \textbf{87}, 174402 (2013).
\bibitem{Zhang2013} L. Zhang, J. Ren, J.-S. Wang, and B. Li, Phys. Rev. B \textbf{87}, 144101 (2013).
\bibitem{Mook2014} A. Mook, J. Henk, and I. Mertig, Phys. Rev. B \textbf{90}, 024412 (2014).
\bibitem{Kimsk2016} S. K. Kim, H. Ochoa, R. Zarzuela, and Y. Tserkovnyak, Phys. Rev. Lett. \textbf{117}, 227201 (2016).
\bibitem{Cheng2016} R. Cheng, S. Okamoto, and D. Xiao, Phys. Rev. Lett. \textbf{117}, 217202 (2016).
\bibitem{Owerre2016} S. A. Owerre, J. Phys.: Condens. Matter \textbf{28}, 386001 (2016).
\bibitem{Zyuzin2016} V. A. Zyuzin and A. A. Kovalev, Phys. Rev. Lett. \textbf{117}, 217203 (2016).
\bibitem{Nakata2017} K. Nakata, S. K. Kim, J. Klinovaja, and D. Loss, Phys. Rev. B \textbf{96}, 224414 (2017).
\bibitem{Gobel2017} B. G\"{o}bel, A. Mook, J. Henk, and I. Mertig, Phys. Rev. B \textbf{95}, 094413 (2017).
\bibitem{Kimsk2019} S. K. Kim, K. Nakata, D. Loss, and Y. Tserkovnyak, Phys. Rev. Lett. \textbf{122}, 057204 (2019).
\bibitem{Diaz2019} S. A. D\'{i}az, J. Klinovaja, and D. Loss, Phys. Rev. Lett. \textbf{122}, 187203 (2019).

\bibitem{Strohm2005} C. Strohm, G. L. J. A. Rikken, and P. Wyder, Phys. Rev. Lett. \textbf{95}, 155901 (2005).
\bibitem{Sheng2006} L. Sheng, D. N. Sheng, and C. S. Ting, Phys. Rev. Lett. \textbf{96}, 155901 (2006).
\bibitem{Inyushkin2007} A. V. Inyushkin and A. N. Taldenkov, JETP Lett. \textbf{86}, 379 (2007).
\bibitem{Kagan2008} Yu. Kagan and L. A. Maksimov, Phys. Rev. Lett. \textbf{100}, 145902 (2008).
\bibitem{LZhang2010} L. Zhang, J. Ren, J.-S. Wang, and B. Li, Phys. Rev. Lett. \textbf{105}, 225901 (2010).
\bibitem{Ideue2017} T. Ideue, T. Kurumaji, S. Ishiwata, and Y. Tokura, Nat. Mater. \textbf{16}, 797 (2017).
\bibitem{Sugii2017} K. Sugii, M. Shimozawa, D. Watanabe, Y. Suzuki, M. Halim, M. Kimata, Y. Matsumoto, S. Nakatsuji, and M. Yamashita, Phys. Rev. Lett. \textbf{118}, 145902 (2017).

\bibitem{Takahashi2016} R. Takahashi, and N. Nagaosa,  Phys. Rev. Lett. \textbf{117}, 217205 (2016).
\bibitem{XZhang2019} X. Zhang, Y. Zhang, S. Okamoto, and D. Xiao, Phys. Rev. Lett. \textbf{123}, 167202 (2019).
\bibitem{Park2019} S. Park and B.-J. Yang,  Phys. Rev. B \textbf{99}, 174435 (2019).
\bibitem{Go2019} G. Go, S. K. Kim, and K.-J. Lee, Phys. Rev. Lett. \textbf{123}, 237207 (2019).
\bibitem{SZhang2020} S. Zhang, G. Go, K.-J. Lee, and S. K. Kim, Phys. Rev. Lett. \textbf{124}, 147204 (2020).


\bibitem{Kittel1949} C. Kittel, Rev. Mod. Phys. \textbf{21}, 541 (1949).

\bibitem{Bazazzadeh2021} N. Bazazzadeh, M. Hamdi, S. Park, A. Khavasi, S. M. Mohseni, and A. Sadeghi, Phys. Rev. B \textbf{104}, L180402 (2021).

\bibitem{McGuire2015} M. A. McGuire, H. Dixit, V. R. Cooper, and B. C. Sales, Chem. Mat. \textbf{27}, 612-620 (2015).
\bibitem{Zhang2015}  W.-B. Zhang, Q. Qu, P. Zhu, and C.-H. Lam, J. Mater. Chem. C \textbf{3}, 12457 (2015).
\bibitem{Lee2016} J. U. Lee, S. Lee, J. H. Ryoo, S. Kang, T. Y. Kim, P. Kim, C.-H. Park, J.-G. Park, and H. Cheong, Nano Lett. \textbf{16}, 7433 (2016).
\bibitem{Gong2017} C. Gong, L. Li, Z. Li, H. Ji, A. Stern, Y. Xia, T. Cao, W. Bao, C. Wang, Y. Wang, Z. Q. Qiu, R. J. Cava, S. G. Louie,
J. Xia, and X. Zhang, Nature (London) \textbf{546}, 265 (2017).
\bibitem{Huang2017} B. Huang, G. Clark, E. Navarro-Moratalla, D. R. Klein, R. Cheng, K. L. Seyler, D. Zhong, E. Schmidgall, M. A. McGuire, D. H. Cobden, W. Yao, D. Xiao,
P. Jarillo-Herrero, and X. Xu, Nature \textbf{546}, 270 (2017).
\bibitem{Bonilla2018} M. Bonilla, S. Kolekar, Y. Ma, H. C. Diaz, V. Kalappattil, R. Das, T. Eggers, H. R. Gutierrez, M.-H. Phan, and M. Batzill, Nat. Nanotechnol. \textbf{13}, 289 (2018).
\bibitem{OHara2018} D. J. O'Hara, T. Zhu, A. H. Trout, A. S. Ahmed, Y. K. Luo, C. H. Lee, M. R. Brenner, S. Rajan, J. A. Gupta, D. W. McComb, and R. K. Kawakami, Nano Lett. \textbf{18}, 3125 (2018).
\bibitem{Fei2018} Z. Fei, B. Huang, P. Malinowski, W. Wang, T. Song, J. Sanchez, W. Yao, D. Xiao, X. Zhu, A. F. May, W. Wu, D. H. Cobden, J.-H. Chu, and X. Xu, Nat. Mater. \textbf{17}, 778 (2018).
\bibitem{Deng2018} Y. Deng, Y. Yu, Y. Song, J. Zhang, N. Z. Wang, Z. Sun, Y. Yi, Y. Z. Wu, S. Wu, J. Zhu, J. Wang, X. H. Chen, and Y. Zhang, Nature (London) \textbf{563}, 94 (2018).
\bibitem{Burch2018} K. S. Burch, D. Mandrus, and J.-G. Park, Nature (London) \textbf{563}, 47 (2018).
\bibitem{Gibertini2019} M. Gibertini, M. Koperski, A. F. Morpurgo, and K. S. Novoselov, Nat. Nanotech, \textbf{14}, 408-419 (2019).


\bibitem{Jiang2018a} S. Jiang, L. Li, Z. Wang, K. F. Mak, and J. Shan, Nat. Nanotech. \textbf{13}, 549 (2018).
\bibitem{Jiang2018b} S. Jiang, J. Shan, and K. F. Mak, Nat. Mater. \textbf{17}, 406 (2018).
\bibitem{Huang2018} B. Huang, G. Clark, D. R. Klein, D. MacNeill, E. Navarro-Moratalla, K. L. Seyler, N. Wilson, M. A. McGuire, D. H. Cobden, D. Xiao, W. Yao, P. Jarillo-Herrero, and X. Xu, Nat. Nanotech. \textbf{13}, 544 (2018).

\bibitem{Kittel1958} C. Kittel, Phys. Rev. 110, 836 (1958).
\bibitem{Thingstad2019} E. Thingstad, A. Kamra, A. Brataas, and A. Sudb{\o}, Phys. Rev. Lett. \textbf{122}, 107201 (2019).
\bibitem{Qu2021} G. Qu, K. Nakamura, and M. Hayashi, J. Phys. Soc. Japan \textbf{90}, 024707 (2021).
\bibitem{Li2020} B. Li, S. Sandhoefner, A. A. Kovalev, Phys. Rev. Research \textbf{2}, 013079 (2020).
\bibitem{Park2020} S. Park and B.-J. Yang, Nano Lett. \textbf{20}, 7694-9699 (2020).
\bibitem{Kovalev2012} A. A. Kovalev and Y. Tserkovnyak, EPL \textbf{97}, 67002 (2012).
\bibitem{Jiang2013} W. Jiang, P. Upadhyaya, Y. Fan, J. Zhao, M. Wang, L.-T.Chang, M. Lang, K. L. Wong, M. Lewis, Y.-T. Lin, J. Tang, S. Cherepov, X. Zhou, Y. Tserkovnyak, R. N. Schwartz,
and K. L. Wang, Phys. Rev. Lett. \textbf{110}, 177202 (2013).
\bibitem{footnote1} The broadening of the phonon band is not important in our calculation because the spin Berry curvatures are induced from the interband transitions between two magnon-like states.

\bibitem{Morell2019} E. S. Morell, A. Leon, R. H. Miwa, and P. Vargas, 2D Mater. \textbf{6}, 025020 (2019).
\bibitem{HZhang2022} H. Zhang and R. Cheng, Appl. Phys. Lett. \textbf{120}, 090502 (2022).
\bibitem{Cheon2018} S. Cheon and H.-W. Lee, Phys. Rev. B \textbf{97}, 104425 (2018).

\end{thebibliography}
\end{document}